# Validation of the CMS Magnetic Field Map

V. I. Klyukhin

*Skobeltsyn Institute of Nuclear Physics, Lomonosov Moscow State University, RU-119992, Moscow, Russia*

Phone : +41-22-767-6561, fax : +41-22-767-7920, e-mail : Vyacheslav.Klyukhin@cern.ch

N. Amapane

*INFN Turin and the University of Turin, I-10125, Turin, Italy*

A. Ball, B. Curé, A. Gaddi, H. Gerwig, M. Mulders

*CERN, CH-1211, Geneva 23, Switzerland*

V. Calvelli

*INFN Genoa and the University of Genoa, I-16146, Genoa, Italy*

A. Hervé, R. Loveless

*University of Wisconsin, WI 53706, Madison, USA*

**Abstract** The Compact Muon Solenoid (CMS) is a general purpose detector, designed to run at the highest luminosity at the CERN Large Hadron Collider (LHC). Its distinctive features include a 4-T superconducting solenoid with 6-m-diameter by 12.5-m-length free bore, enclosed inside a 10,000-ton return yoke made of construction steel. The return yoke consists of five dodecagonal three-layered barrel wheels and four end-cap disks at each end comprised of steel blocks up to 620 mm thick, which serve as the absorber plates of the muon detection system. To measure the field in and around the steel, a system of 22 flux loops and 82 3-D Hall sensors is installed on the return yoke blocks. A TOSCA 3-D model of the CMS magnet is developed to describe the magnetic field everywhere outside the tracking volume measured with the field-mapping machine. The magnetic field description is compared with the measurements and discussed.

**Keywords** Flux loops, Hall probes, Magnetic field measurements, Superconducting solenoid

## 1 Introduction

The muon system of the Compact Muon Solenoid (CMS) detector includes a 10,000-ton yoke comprised of the construction steel plates up to 620 mm thick, which return the flux of the 4 T superconducting solenoid and serve as the absorber plates of the muon detection system [1–4]. During the LHC long shutdown occurring in 2013/2014 the CMS magnet yoke is upgrading with the additional 14-m-diameter end-cap disks at the extremes of the muon detection system [5]. The presence of the 0.125-m-thick disks changes the magnetic flux density distribution in the adjacent end-cap disks by 25% in average. This requires developing the new magnetic field map to be used in the detector simulation and the event reconstruction software.

The magnetic flux density in the central part of the CMS detector, where the tracker and electromagnetic calorimeter are located, was measured with precision of $7 \cdot 10^{-4}$ with the field-mapping machine at five central field values of 2, 3, 3.5, 3.8, and 4 T [6]. To describe the magnetic flux everywhere outside the measured volume, a three-dimensional (3-D) magnetic field model of the CMS magnet has been developed [7] and calculated with TOSCA [8] when the detector began operation. The model reproduces the magnetic flux density distribution measured inside the CMS coil with the field-mapping machine within 0.1% [9]. The modification of this model for the upgraded CMS magnet yoke requires validating the model by comparing the calculated magnetic flux density with the measured one at least in selected regions of the CMS magnetic system.

A direct measurement of the magnetic flux density in the yoke-selected regions was provided during the CMS magnet test of 2006 with 22 flux loops of 315÷450 turns wound around the yoke blocks. The "fast" (190 s time-constant) discharges of the CMS coil made possible by the protection system, which is provided to protect the magnet in the event of major faults [10, 11], induced in the flux loops the voltages caused by the magnetic flux changes. An integration technique [12, 13] was developed to reconstruct the average initial magnetic flux density in steel blocks at the full magnet excitation, and the contribution of the eddy currents was calculated with ELECTRA [14] and estimated on the level of a few per cent [15].

The results of the magnetic flux measurements done with the flux loops and comparison the obtained values with the calculations performed with the previous TOSCA CMS magnet model are described elsewhere [16].

In present paper we compare the calculations done with the recent CMS magnet model with the measurements performed with the flux loops in the yoke steel and with the 3-D Hall probes installed at the steel-air interfaces in the gaps between the CMS yoke parts. These comparisons allow validating the magnetic field maps to be used for the upgraded CMS detector.

## 2 The CMS Magnet Model Description

The CMS magnet model for the upgraded detector is presented in Fig. 1. The central part of the model comprises the coil four layers of the superconductor and the full magnet yoke that consists of five barrel wheels of the 13.99 m inscribed outer diameter and 2.536 m width, two nose disks of 5.26 m diameter on each side of the coil, three large end-cap disks of the 13.91 m inscribed outer diameter on each side of the magnet, two small end-cap disks of 5 m diameter, and two additional 0.125-m-thick end-cap disks of the 13.91 m inscribed outer diameter mounted around these small disks.

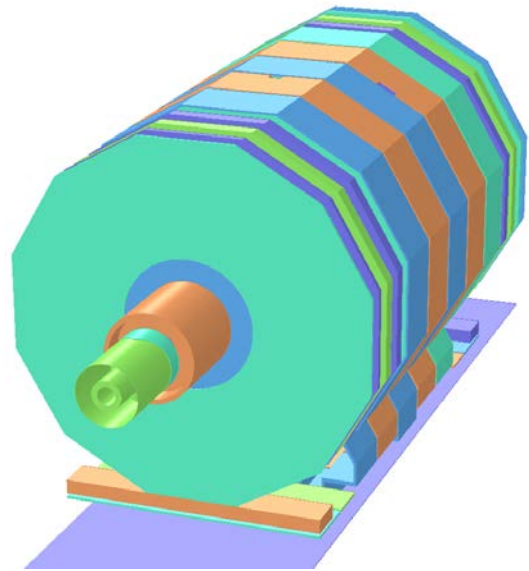

**Fig. 1** 3-D model of the CMS magnetic system

The total length of the central part is 21.61 m. On both sides of the central part the hadronic forward calorimeter absorbers and shields, the steel collars and rotating shields of the beam pipe are included into the model. The 40-mm-thick steel floor of the experimental underground cavern has the length of 36.4 m and the width of 9.9 m.

Each barrel wheel except of central one has three layers of steel connected with brackets. The central barrel wheel comprises the fourth most inner layer, tail catcher, made of steel and turned by 5º in the azimuth angle with respect to dodecagonal shape of the barrel wheels. The coordinate system used in the model corresponds to the CMS reference system where the *X*-axis is directed in horizontal plane toward the LHC center, the *Y*-axis is upward, and the *Z*-axis coincides with the superconducting coil axis and has the same direction as the positive axial component of the magnetic flux density.

The model comprises 21 conductors, the barrel feet and the end-cap disk carts and contains 7,111,713 nodes of the finite element mesh shown in the *XY*-plane in Fig. 2.

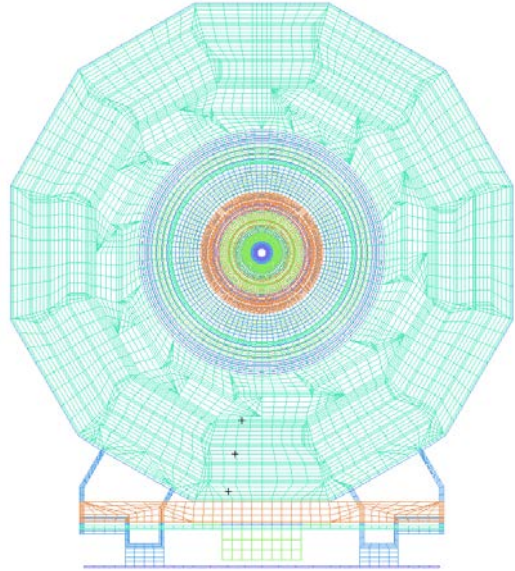

**Fig. 2** The *XY*-plane of the finite element mesh used to extrude the 3-D model. The *black crosses* near the azimuth angle of 255° display the location of the 3-D Hall sensors measured the magnetic flus density at the steel-air interfaces

The dimensions of the yoke parts and the superconducting coil modules are described elsewhere [16]. The operational current of the CMS superconducting coil is 18.164 kA.

## 2.1 Steel Magnetic Properties Description

Three different *B-H* curves of the construction steel of the CMS magnet yoke are used in the model. The first curve describes the magnetic properties of the barrel wheel thick plates in the second and third layers. Second curve describes the magnetic properties of thin plates around the thick plates of the second and third barrel wheel layers, and also the properties of the plates of the first layers and the tail catcher plates of the barrel wheels. This curve is used as well for the extensions of the forth end-cap disks, the forward hadronic calorimeter absorbers and shields, and steel collars and rotating shields around the beam pipe. Finally, the third curve describes the magnetic properties of the nose and end-cap disks, the cart plates, keels, and the steel floor.

# 3 Comparison of the Measured and Calculated Magnetic Flux Density

The measurements used for the comparisons were obtained in the CMS magnet test during two current cycles on August 17 and 28, 2006 shown in Fig. 3.

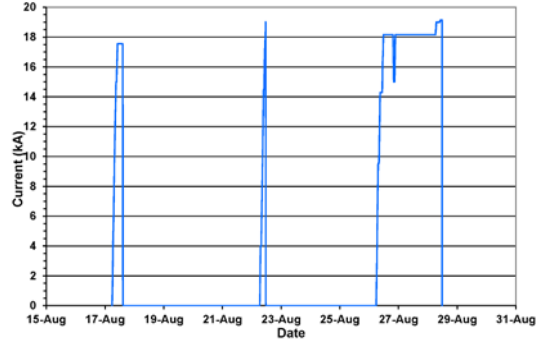

**Fig. 3** The CMS magnet current cycles of August 2006 used in the flux loops and 3-D Hall sensors measurements at 17.55 kA (17-Aug) and 19.14 kA (28-Aug)

In the first cycle on August 17 the current reaches the maximum value of 17.55 kA and after 140 min the fast discharge was manually triggered. In the second cycle on August 28 the current reaches the maximum value of 19.14 kA and after 71 min the fast discharge was triggered accidently. These flat current time intervals were used to measure the magnetic flux density with the 3-D Hall sensors installed in the *XY*-planes at *Z*-coordinates of 1.273, –1.418, –3.964, –4.079, –6.625, and –7.251 m. In present analysis only 18 sensors located in the plane tilted at $X < 0$ m by the angle of 15° with respect to the vertical plane and shifted at the *Y*-coordinate of –4.805 m with respect to the vertical plane by 0.56 m as shown in Fig. 2 are used.

The magnetic flux density in the CMS yoke steel is measured using the voltages induced in the flux loops during the fast discharges of the coil from the maximum currents. The average magnetic flux density in steel was reconstructed by the off-line integration of the voltages [16]. The current values of 17.55 and 19.14 kA produce the magnetic flux density in the CMS coil center of 3.68, and 4.01 T, accordingly. The operational central magnetic flux density of 3.81 T used in the CMS magnet is between these values.

In Figs. 4–5 the measured values of the magnetic flux density vs. *Z*- and *Y*-coordinates are displayed and compared with the calculated field values obtained with new CMS model at the currents of 17.55 and 19.14 kA. To perform the comparisons, the model geometry of the same finite element mesh is adapted to the configuration of the CMS magnet used in the 2006 test. The extensions of the forth end-cap disks, one small forth end-cap disk, the steel floor, the forward hadronic calorimeter absorbers and shields, steel collars and rotating shields are removed from the model. The contribution of all the removed ferromagnetic parts into the central magnetic flux density is 0.03%. The same contribution of the yoke central part is 7.97%. Thus, the configuration of the yoke used in the 2006 test is acceptable for the model validation.

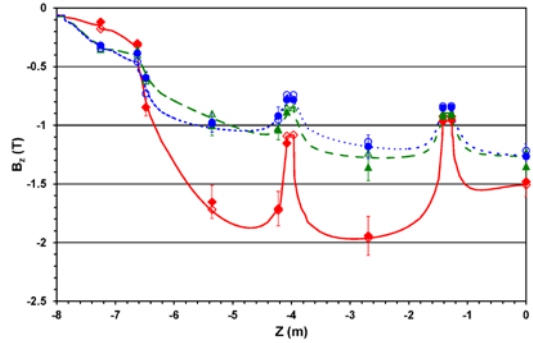

**Fig. 4** Axial magnetic flux density measured (*filled markers*) and calculated (*opened markers*) in the first (*rhombs*), second (*triangles*), and third (*circles*) barrel layers vs. Z-coordinate at the current of 17.55 kA. The *lines* represent the calculated values along the Hall sensors locations at the Y-coordinates of –4.805 m (*solid line*), –5.66 m (*dashed line*), and –6.685 m (*dotted line*), corresponded to the first, second, and third barrel layer blocks, respectively

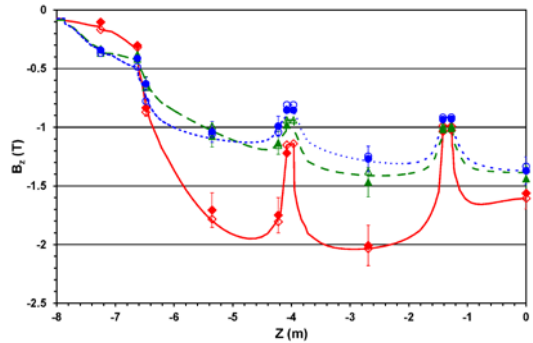

**Fig. 5** Axial magnetic flux density measured (*filled markers*) and calculated (*opened markers*) in the first (*rhombs*), second (*triangles*), and third (*circles*) barrel layers vs. Z-coordinate at the current of 19.14 kA. The *lines* represent the calculated values along the Hall sensors locations at the *Y*-coordinates of –4.805 m (*solid line*), –5.66 m (*dashed line*), and –6.685 m (*dotted line*), corresponded to the first, second, and third barrel layer blocks, respectively

The comparison gives the differences between the calculated and measured values of the magnetic flux density as follows: (0.59 ± 7.41) % in the barrel wheels and (–4.05 ± 1.97) % in the end-cap disks at the maximum current of 17.55 kA; (1.41 ± 7.15) % in the barrel wheels and (–2.87 ± 2.00) % in the end-cap disks at the maximum current of 19.14 kA. The error bars of the magnetic flux density measured with the flux loops are of ±8.55% and include the errors in the knowledge of the flux loops geometries and the errors of the measured magnetic fluxes. The error bars of the 3-D Hall sensor measurements are ±(0.025 ± 0.015) mT at the current of 17.55 kA and ±(0.012 ± 0.001) mT at the current of 19.14 kA. In addition to these comparisons the model perfectly describes the magnetic flux density distribution inside the CMS coil within 0.1% in accordance with the previous model consisted of two separate halves of the yoke [9], [16].

## 4 Conclusions

The new CMS magnet model is developed to prepare the magnetic field maps for the upgraded CMS detector. The model is validated by the comparison of the calculated magnetic flux density with the measurements done in the CMS magnet selected regions with the flux loops and 3-D Hall sensors.